\documentclass[aps,pre,amsmath,amssymb,reprint,superscriptaddress]{revtex4-1}
\usepackage[dvipdfmx]{graphicx}
\usepackage{bmpsize}
\usepackage{txfonts}
\usepackage{color}
\usepackage{epsf}
\usepackage{bm}
\usepackage{lipsum}

\begin{document}

\title{Understanding the scaling of boson peak through insensitivity of elastic heterogeneity to bending rigidity in polymer glasses}

\author{Naoya Tomoshige}
\thanks{Equally contributed to this work.}
\affiliation{Division of Chemical Engineering, Department of Materials Engineering Science, Graduate School of Engineering Science, Osaka University, Toyonaka, Osaka 560-8531, Japan}

\author{Shota Goto}
\thanks{Equally contributed to this work.}
\affiliation{Division of Chemical Engineering, Department of Materials Engineering Science, Graduate School of Engineering Science, Osaka University, Toyonaka, Osaka 560-8531, Japan}

\author{Hideyuki Mizuno}
\email{hideyuki.mizuno@phys.c.u-tokyo.ac.jp}
\affiliation{Graduate School of Arts and Sciences, The University of Tokyo, Tokyo 153-8902, Japan}

\author{Tatsuya Mori}
\affiliation{Department of Materials Science, University of Tsukuba, 1-1-1 Tennodai, Tsukuba, Ibaraki, 305-8573, Japan}

\author{Kang Kim}
\email{kk@cheng.es.osaka-u.ac.jp}
\affiliation{Division of Chemical Engineering, Department of Materials Engineering Science, Graduate School of Engineering Science, Osaka University, Toyonaka, Osaka 560-8531, Japan}

\author{Nobuyuki Matubayasi}
\email{nobuyuki@cheng.es.osaka-u.ac.jp}
\affiliation{Division of Chemical Engineering, Department of Materials Engineering Science, Graduate School of Engineering Science, Osaka University, Toyonaka, Osaka 560-8531, Japan}

\date{\today}

\begin{abstract}
Amorphous materials exhibit peculiar mechanical and vibrational
 properties, including non-affine elastic responses and excess
 vibrational states, \textit{i.e.}, the so-called boson peak.
For polymer glasses, these properties are considered to be affected by
 the 
 bending rigidity of the constituent polymer chains.
In our recent work~[Tomoshige, \textit{et al.}, Sci. Rep. \textbf{9}
 19514 (2019)], we have revealed simple 
relationships between the variations of vibrational properties and the
 global elastic properties: the response of the boson peak scales only with 
 that of the global shear modulus.
This observation suggests that the spatial
 heterogeneity of the local shear modulus distribution is insensitive to 
 changes in the bending rigidity.
Here, we demonstrate the insensitivity of elastic heterogeneity by directly measuring
 the local shear modulus distribution.
We also study transverse sound wave propagation, which is also shown to
 scale only with 
 the global shear modulus.
Through these analyses, we conclude that the bending rigidity does not
 alter the spatial heterogeneity of the local shear modulus distribution,
 which yields vibrational and acoustic properties 
 that are controlled solely by the global shear modulus of a polymer glass.
\end{abstract}

\maketitle

\section{Introduction}
Amorphous materials exhibit anomalous mechanical and vibrational
properties that have been studied for many years by experimental, 
numerically, and theoretical methods.
The vibrational and acoustical properties of such materials have been
investigated 
in many experiments using neutron, light, and X-ray scattering,
\textit{e.g.},
Refs.~\cite{Monaco:2006kz, Baldi:2010bk, Chumakov:2011hl, Yamamuro:1996dd,
Ramos:2003hn, Monaco:2009eg, Shibata:2015ei, Kabeya:2016eb, Mori:2020ci}.  
Using these methods, 
anomalies in vibrational and acoustic
excitations have been detected, including excess vibrational states, the
so-called boson
peak (BP), and strong damping of sound wave propagation.

To explain these anomalous properties, the heterogeneous elasticity
theory was proposed and developed by Schirmacher and
co-workers~\cite{Schirmacher:2006ky, Schirmacher:2007cj, Kohler:2013jj,
Schirmacher:2015bq} (see
also Refs.~\cite{Wyart:2010dw, DeGiuli:2014hu} for the theory in the
context of the jamming transition and
Refs.~\cite{Shimada:2020ka, Shimada:2020vi, Shimada:2020vc} for very recent
developments).
It is now well-established that amorphous materials exhibit spatial
heterogeneity in their local elastic modulus distributions, as supported by numerical
simulations~\cite{Tsamados:2009ke, Makke:2011cv, Mizuno:2013fc} and 
experiments~\cite{Wagner:2011ck, Hufnagel:2015bl}.
In the theory, elastic moduli heterogeneities are critical in
describing 
anomalies in the vibrational, acoustic, and thermal properties. 
The theory notably predicts that the BP and the attenuation rate
of sound are more significant when moduli distributions are more
heterogeneous.
This prediction has been tested and justified by numerical
simulations~\cite{Marruzzo:2013ew, Marruzzo:2013bw, Mizuno:2013gu,
Mizuno:2014dz, Mizuno:2016cv, Shakerpoor:2020hd, Kapteijns:2021jj}.

Anomalous behaviours in polymer glasses have also been reported through
both experiments~\cite{Niss:2007de, Hong:2008ix, Caponi:2011ip,
PerezCastaneda:2014bx, Terao:2018kw, Zorn:2018ed, Corezzi:2020ha}
and numerical
simulations~\cite{Jain:2004kj, Schnell:2011he, Lappala:2016dn,
Ness:2017cc, Milkus:2018ha, Giuntoli:2018fv}.
In polymer glasses, the bending rigidity
of the constituent polymer chains is an important parameter.
In our recent work~\cite{Tomoshige:2019jm}, we studied the effects of
the bending rigidity on the global elastic moduli (shear modulus $G$ and bulk
modulus $K$) and the vibrational density of states (vDOS) $g(\omega)$ using
coarse-grained molecular dynamics (MD) simulations.
We demonstrated 
that the variation of the BP simply follows that of global shear modulus
$G$ through the Debye frequency $\omega_\mathrm{D}$.
If this simple scaling behaviour is considered in terms of the heterogeneous
elasticity theory,
we obtain an important implication that the spatial heterogeneity in
local modulus distributions is insensitive to changes in the bending rigidity.

In this study, we examine this correlation by 
directly measuring the degree of
elastic heterogeneity with changes in the bending rigidity.
We also study transverse acoustic excitations in the polymer glasses by calculating
the dynamic structure factor and examine the connection among 
the sound velocity, attenuation rate, and the simple scaling
behaviour of the BP.
Thus, we comprehensively discuss that the effects of bending rigidity in
polymer glasses on 
vibrational and acoustic excitations from the perspective of elastic heterogeneities.

The remainder of this paper is organized as follows. 
Section~\ref{sec:simulation} describes the MD simulation details used to
characterise the elastic heterogeneity and the acoustic excitation.
In Section~\ref{sec:result}, the numerical results and discussions are presented.
Finally, our conclusions are drawn in Section~\ref{sec:conclusion}.

\section{Simulation method}
\label{sec:simulation}
%
\subsection{Simulation model}
We performed MD simulations using the Kremer--Grest
model~\cite{Kremer:1990iv}, which is a coarse-grained bead-spring
model of the polymer chain.
Each polymer chain comprises $L$ monomer beads of mass $m$ and
diameter $\sigma$.
We studied the case of $200$ chains of $L=50$, such that the system
contained $N = 200 \times 50 = 10000$ monomer beads in total, in a
three-dimensional cubic box of volume $V$ under periodic boundary conditions.

In the Kremer--Grest model, three types of inter-particle potentials are utilised.
First, the Lennard-Jones (LJ) potential
\begin{equation}
U_\mathrm{LJ}(r) = 4\varepsilon_\mathrm{LJ} \left[ \left( \frac{\sigma}{r} \right)^{12} - \left( \frac{\sigma}{r} \right)^{6} \right],
\label{eq:LJ}
\end{equation}
acts between all pairs of monomer beads, where $r$ and
$\varepsilon_\mathrm{LJ}$ denote the distance between two monomers and
the energy scale of the LJ potential, respectively.
The LJ potential is truncated at the cut-off distance of $r_c =
2.5\sigma$, where the potential and the force (the first derivative of the
potential) are shifted to zero continuously~\cite{Shimada:2018fp}.

Second, sequential monomer beads along the polymer chain are connected
by a finitely extensible nonlinear elastic (FENE) potential:
\begin{equation}
U_\mathrm{FENE}(r) =
\begin{cases}
-\frac{\varepsilon_\mathrm{FENE}}{2} R_0^2 \ln \left[ 1 -
 \left(\frac{r}{R_0} \right)^{2} \right]  \quad &(r \le R_0),\\
\infty  \quad &(r >R_0),
\end{cases}
\label{eq:FENE}
\end{equation}
where $\varepsilon_\mathrm{FENE}$ is the energy scale of the FENE
potential, and $R_0$ is the maximum length of the FENE bond.
Following Ref.~\cite{Milkus:2018ha}, we employ the values of
$\varepsilon_\mathrm{FENE} = 30 \varepsilon_\mathrm{LJ}$ and $R_0 = 1.5
\sigma$.

Finally, the bending angle $\theta$ formed by three consecutive monomer
beads along the polymer chain is controlled by
\begin{equation}
U_\mathrm{bend}(\theta) = \varepsilon_\mathrm{bend} \left[ 1 - \cos(\theta - \theta_0) \right],
\label{eq:bend}
\end{equation}
where $\varepsilon_\mathrm{bend}$ is the associated bending energy.
We set the stabilised angle as $\theta_0 = 109.5^\circ$~\cite{Milkus:2018ha}.
In the present work, we utilise a wide range of $\varepsilon_\mathrm{bend}$ values:
$\varepsilon_\mathrm{bend}/\varepsilon_\mathrm{LJ} = 10^{-1}$, $1$, $3$,
$10$, $30$, $10^{2}$, $3\times 10^{2}$, $10^3$, and $3\times 10^3$.

We performed the MD simulations using the Large-scale Atomic/Molecular
Massively Parallel Simulator (LAMMPS)~\cite{Plimpton:1995wl}.
Hereafter, the length, energy, and time are measured in units of
$\sigma$, $\varepsilon_\mathrm{LJ}$, and
$\sigma(m/\varepsilon_\mathrm{LJ})^{1/2}$, respectively.
The temperature is presented in units of
$\varepsilon_\mathrm{LJ}/k_\mathrm{B}$, where $k_\mathrm{B}$ is the
Boltzmann constant. 
We first equilibrated the polymer melt system at a temperature $T=1.0$
and polymer bead number density $\hat\rho =N/V=0.85$.
We then cooled the system down towards $T=0.05$ with a constant cooling
rate of $dT/dt = 10^{-4}$, under a fixed pressure of $p =0$.
Finally, the inherent structure at $T = 0$ is generated using the
steepest descent method.
In our recent work~\cite{Tomoshige:2019jm}, we reported
the dependence of the glass transition
temperature $T_g$ and the number density $\hat\rho$ at zero temperature
on $\varepsilon_\mathrm{bend}$.

\subsection{Vibrational density of state and boson peak}

The vDOS analysis was performed for the configuration 
at $T=0$.
By diagonalizing the Hessian matrix, we obtained the eigenvalues
$\lambda^k$ ($k=1$, 2, $\cdots$, $3N$), which provide the
eigenfrequencies as $\omega^k = \sqrt{\lambda^k}$.
The vDOS is defined as
\begin{equation}
g(\omega) = \frac{1}{3N-3} \sum_{k=1}^{3N-3} \delta(\omega-\omega_k), 
\end{equation}
where three zero-frequency modes are omitted.
The expression of the Hessian matrix of the polymeric system was given in Ref.~\cite{Tomoshige:2019jm}.
The Debye law predicts the vDOS as
$g_\mathrm{D}(\omega)=\omega^2 A_\mathrm{D}$, where 
$A_\mathrm{D} = 3/{\omega_\mathrm{D}}^3$ is the Debye level using the Debye frequency
$\omega_\mathrm{D}=[18\pi^2\rho/(2{c_\mathrm{T}}^{-3}+{c_\mathrm{L}}^{-3})]^{1/3}$.
Here, the transverse and longitudinal sound velocities,
$c_\mathrm{T}$ and  $c_\mathrm{L}$, are given by the bulk modulus $K$,
shear modulus $G$, and the mass density $\rho=m\hat\rho$ as 
$c_\mathrm{T}=\sqrt{G/\rho}$ and $c_\mathrm{L} =\sqrt{(K+4G/3)/\rho}$, respectively.
The reduced vDOS $g(\omega)/\omega^2$ thus characterises the excess
vibrational modes exceeding the Debye prediction,
\textit{i.e.}, the BP.

\subsection{Global and local shear modulus}
The global shear modulus $G$ and bulk modulus $K$ were evaluated from the
stress-tensor response to the shear and volume deformations in the
``quasi-static'' way, respectively, applied to the inherent structure.
For perfect crystalline solids, the mechanical equilibrium is maintained
during affine deformation.
However, the force balance is generally broken down for amorphous solids
under applied affine deformations.
Thus, further energy minimization causes additional non-affine
deformation (relaxation) towards mechanical equilibrium.
In other words, $G$ and $K$ are decomposed into
$G=G_\mathrm{A}-G_\mathrm{NA}$ and $K=K_\mathrm{A}-K_\mathrm{NA}$.
Here, $M_\mathrm{A}$ and $M_\mathrm{NA}$ denote the affine and
non-affine components of elastic moduli, with $M=G$ and $K$,
respectively.
Our recent work~\cite{Tomoshige:2019jm} also reported the
$\varepsilon_\mathrm{bend}$ dependence of $G$ and $K$.
In particular, we demonstrated that the bulk modulus $K$ is much
larger than the shear modulus $G$, and thus the shear modulus has 
important effects on the low-frequency vibrational properties of the
polymeric system.

In this study, we further study the \textit{local} shear modulus.
Specifically, we measure the spatial distribution of the local shear
modulus $G_\mathrm{m}$, by using the numerical procedure of ``affine
strain approach'', given in Ref.~\cite{Mizuno:2013fc}.
Note that the analysis completely neglects anharmonic effects
and provide zero-temperature limit values of elastic heterogeneities.
Briefly, we divided the system into 7$\times$7$\times 7$ cubic
cells and monitored the local shear stress as a function of the applied
shear strain in each local cell.
The linear dimension of the cell is approximately $W\approx 3\sigma$.
Here, the local strain of the small cell is assumed to be
given by the global strain applied to the system.
The local shear modulus $G_m$ of cell $m$ was
measured as the slope of the local shear stress versus the shear strain.
The expression of the local modulus was also given in
Ref.~\cite{Mizuno:2013fc}.
Finally, we collected the $G_m$ values for all the cells to calculate the
probability distribution of the local shear modulus $P(G_\mathrm{m})$.
Remark that the average and standard deviation of
the local shear modulus distribution is insensitive to the cell size $W$~\cite{Mizuno:2013fc}.

As in the LJ glass~\cite{Tsamados:2009ke, Mizuno:2013fc},
we found that $P(G_m)$ is well fitted to the Gaussian 
\begin{equation}
P(G_\mathrm{m})=\frac{1}{\sqrt{2\pi}\delta G_\mathrm{m}}
 \exp{\left[-\frac{(G_\mathrm{m}-G)^2}{2{\delta
       G_\mathrm{m}}^2}\right]},
\label{eq:Gaussian}
\end{equation}
where the relative standard deviation $\delta G_\mathrm{m}/G$ provides a
measure of the spatial heterogeneity in the local shear modulus distribution.

\subsection{Transverse acoustic excitation}
The transverse acoustic excitation can be characterised by 
the (transverse) dynamic structure factor as a function of the wave vector
$\bm{q}$ and frequency $\omega$~\cite{Marruzzo:2013ew, Mizuno:2014dz,
Monaco:2009da, Beltukov:2016en}:
\begin{equation}
S_\mathrm{T}(q,\omega) = \left(\frac{q}{\omega}\right)^2 \frac{1}{2\pi}
 \int\frac{1}{N}\langle
 \bm{j}_\mathrm{T}(\bm{q},t)\cdot\bm{j}_\mathrm{T}^{*}(\bm{q},0)
 \rangle\exp(i\omega t)dt,
\label{eq:DSF}
\end{equation}
where $q=|\bm{q}|$, `$\ast$' indicates complex conjugation, and
$\langle \cdots \rangle$ denotes the ensemble average over the initial time and
angular components of $\bm{q}$.
Here, the transverse current is expressed by:
\begin{equation}
  \bm{j}_\mathrm{T}(\bm{q},t) = \sum_{i=1}^{N}[\bm{v}_i(t)-(\bm{v}_i(t)\cdot\hat{\bm{q}})\hat{\bm{q}}]\exp\left[ i\bm{q}\cdot\bm{r}_i(t) \right],
\end{equation}
where $\hat{\bm{q}}=\bm{q}/q$, and $\bm{r}_i$ and $\bm{v}_i
(=d\bm{r}_i/dt)$ represent the position and velocity, respectively, of
the monomer bead $i$.
In general, the dynamic structure factor $S(q, \omega)$ exhibits two
kinds of peaks: the Rayleigh (elastic) peak and the
Brillouin (inelastic) peak.
The Rayleigh peak is located at $\omega\to 0$ and is related to the
thermal diffusion, while the Brillouin peak is related to
the (transverse) sound-wave propagation.

The Brillouin peak in $S_\mathrm{T}(q, \omega)$ can be fitted by the
damped harmonic oscillator function~\cite{Marruzzo:2013ew,
Mizuno:2014dz, Monaco:2009da, Beltukov:2016en},
\begin{equation}
S_\mathrm{T}(q, \omega) \propto \frac
 {\Gamma_\mathrm{T}(q)\Omega_\mathrm{T}^2(q)}
 {[\omega^2-\Omega_\mathrm{T}^2(q)]^2+\omega^2\Gamma_\mathrm{T}^2(q)},
\label{eq:DHO}
\end{equation}
which provides information about the propagation frequency
$\Omega_\mathrm{T}(q)$ and the attenuation rate $\Gamma_\mathrm{T}(q)$
as functions of the wave number $q$.
The sound velocity is then given by $c_\mathrm{T}(q) =
\Omega_\mathrm{T}(q)/q$.
Note that the sound velocity $c_\mathrm{T}(q)$ converges to the
macroscopic value $c_\mathrm{T}=\sqrt{G/\rho}$ in the long-wavelength limit of $q \to 0$.
We numerically calculated the dynamic structure factor $S_\mathrm{T}(q,\omega)$
[Eq.~(\ref{eq:DSF})] of the inherent structure for each bending energy
$\varepsilon_\mathrm{bend}$ from $\varepsilon_\mathrm{bend} = 10^{-1}$
to $3\times 10^3$.
Note that the thermal fluctuations are imposed at very low temperature
$T=0.05$, which is  small enough that the derived
values are consistent with the zero-temperature limit values.
The values of $\Omega_\mathrm{T}(q)$ and $\Gamma_\mathrm{T}(q)$ were
then extracted using Eq.~(\ref{eq:DHO}).

\section{Results and Discussion}
\label{sec:result}

\begin{figure}[t]
\centering
\includegraphics[width=0.48\textwidth]{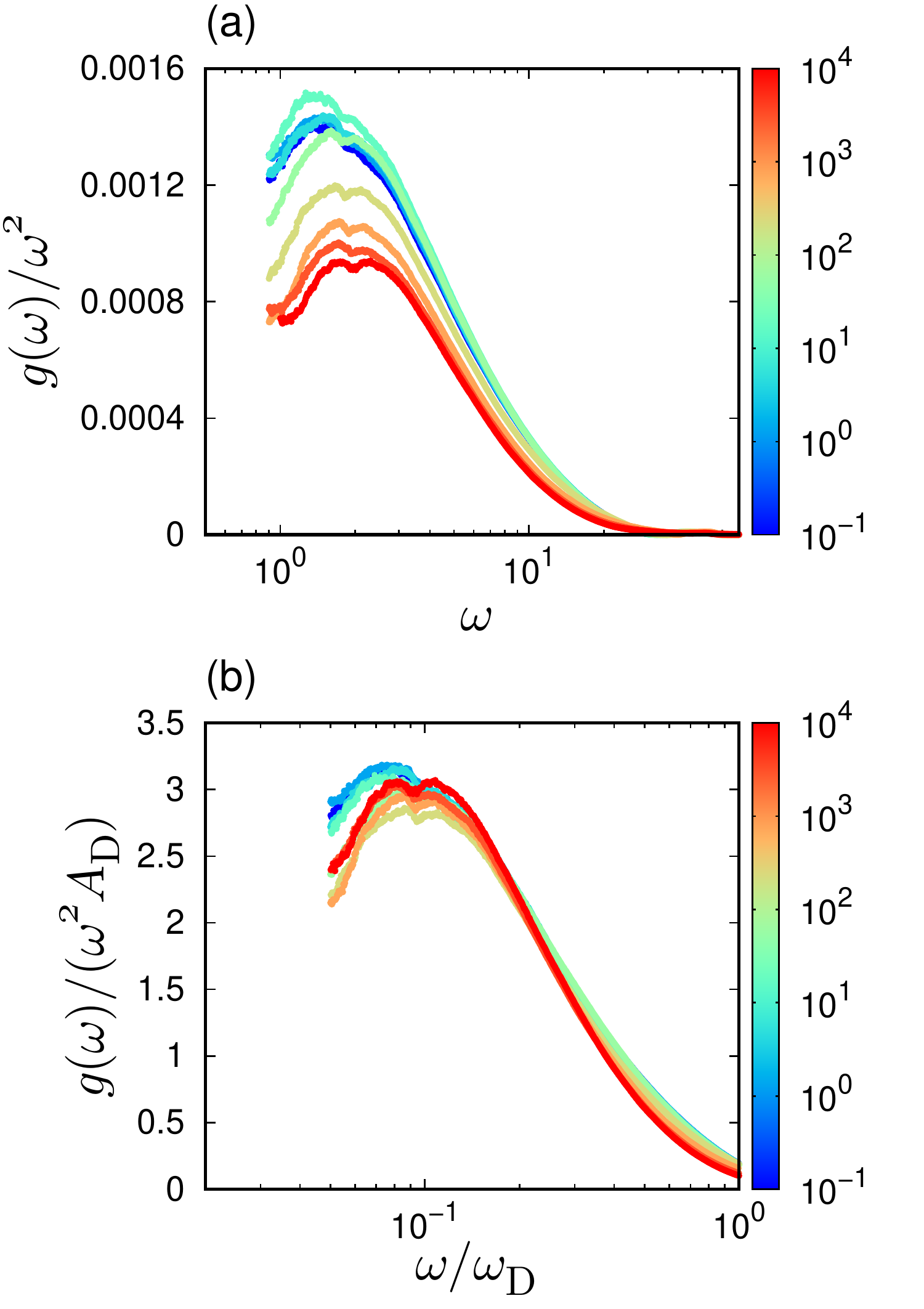}
\vspace*{0mm}
\caption{\label{fig1}
(a) The reduced vDOS $g(\omega)/\omega^2$ with changing the strength of bending rigidity $\varepsilon_\mathrm{bend}$.
(b) The reduced vDOS $g(\omega)/(\omega^2 A_\mathrm{D})$ scaled by the Debye level
 $A_\mathrm{D}$ as a function of the frequency
 $\omega/\omega_\mathrm{D}$ scaled further by the Debye frequency $\omega_\mathrm{D}$.
The color of line indicates the value of bending rigidity $\varepsilon_\mathrm{bend}$ according to the color bar.
}
\end{figure}

\begin{figure}[t]
\centering
\includegraphics[width=0.48\textwidth]{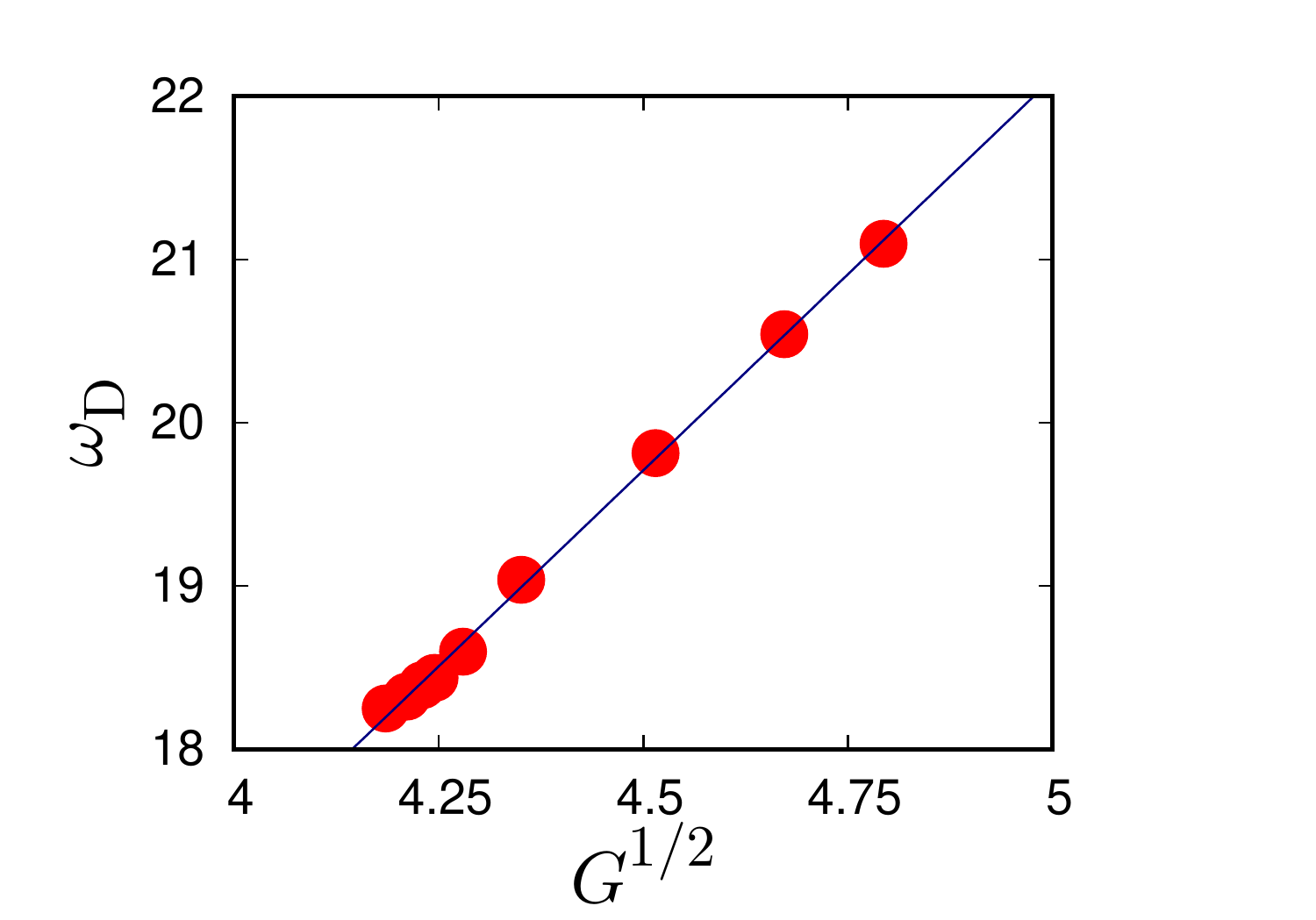}
\vspace*{0mm}
\caption{\label{fig2}
Debye frequency $\omega_\mathrm{D}$ versus square root of global shear modulus $G$.
The straight line is a viewing guide for $\omega_\mathrm{D} \propto G^{1/2}$.
From left to right, the bending energy changes from $\varepsilon_\mathrm{bend} = 10^{-1}$ to $3\times 10^3$.
}
\end{figure}

\subsection{Scaling of boson peak by the Debye frequency and Debye level}
Figure~\ref{fig1}(a) plots the reduced vDOS $g(\omega)/\omega^2$, showing
the BP beyond the Debey level $A_\mathrm{D}$ for each
$\varepsilon_\mathrm{bend}$.
The BP frequency $\omega_\mathrm{BP}$ is located at $\omega_\mathrm{BP}
\approx 2$, but it slightly shifts to the higher frequency with increasing the
bending rigidity.
In addition, the peak height of $g(\omega)/\omega^2$ gradually decreases
when $\varepsilon_\mathrm{bend}$ is increased.
Figure~\ref{fig1}(b) shows 
the reduced vDOS $g(\omega)/(\omega^2 A_\mathrm{D})$ scaled by the Debye
level $A_\mathrm{D}$ as a function of the frequency $\omega/\omega_\mathrm{D}$ scaled
by the Debye frequency $\omega_\mathrm{D}$.
This demonstrates the scaling of the BP by the Debye frequency
$\omega_\mathrm{D}$ and Debye level $A_\mathrm{D}$ for various bendig rigidities of
the polymer chain.
Note that the scaling property of the BP is also shown for
shorter polymer chain with the length $L=3$ in our previous paper~\cite{Tomoshige:2019jm}.

\subsection{Debye frequency and global shear modulus}
We next examine the relationship between the Debye frequency
$\omega_\mathrm{D}$ and the shear modulus $G$, which is plotted in Fig.~\ref{fig2}.
As demonstrated in Ref.~\cite{Tomoshige:2019jm}, the bulk modulus $K$ is
approximately three to four times larger than the shear modulus $G$.
Thus the term ${c_\mathrm{L}}^{-3}$ becomes negligible, and 
the Debye frequency $\omega_\mathrm{D}$ can be approximated as
\begin{equation}
\omega_\mathrm{D} = \left[
		     \frac{18\pi^2\rho}{2{c_\mathrm{T}}^{-3}+{c_\mathrm{L}}^{-3}}
		    \right]^{1/3} \simeq (9\pi^2\rho)^{1/3} c_\mathrm{T}
\propto \sqrt{G},
\label{eq:Debye_frequency}
\end{equation}
which is mainly governed by the shear modulus $G$.
Figure~\ref{fig2} directly demonstrates the relationship of
$\omega_\mathrm{D}\propto \sqrt{G}$ with changes in $\varepsilon_\mathrm{bend}$.
The density $\rho$ is also changed by changing
$\varepsilon_\mathrm{bend}$, but the effect of density on
$\omega_\mathrm{D}$ is close to negligible~\cite{Tomoshige:2019jm}.

\begin{figure}[t]
\centering
\includegraphics[width=0.48\textwidth]{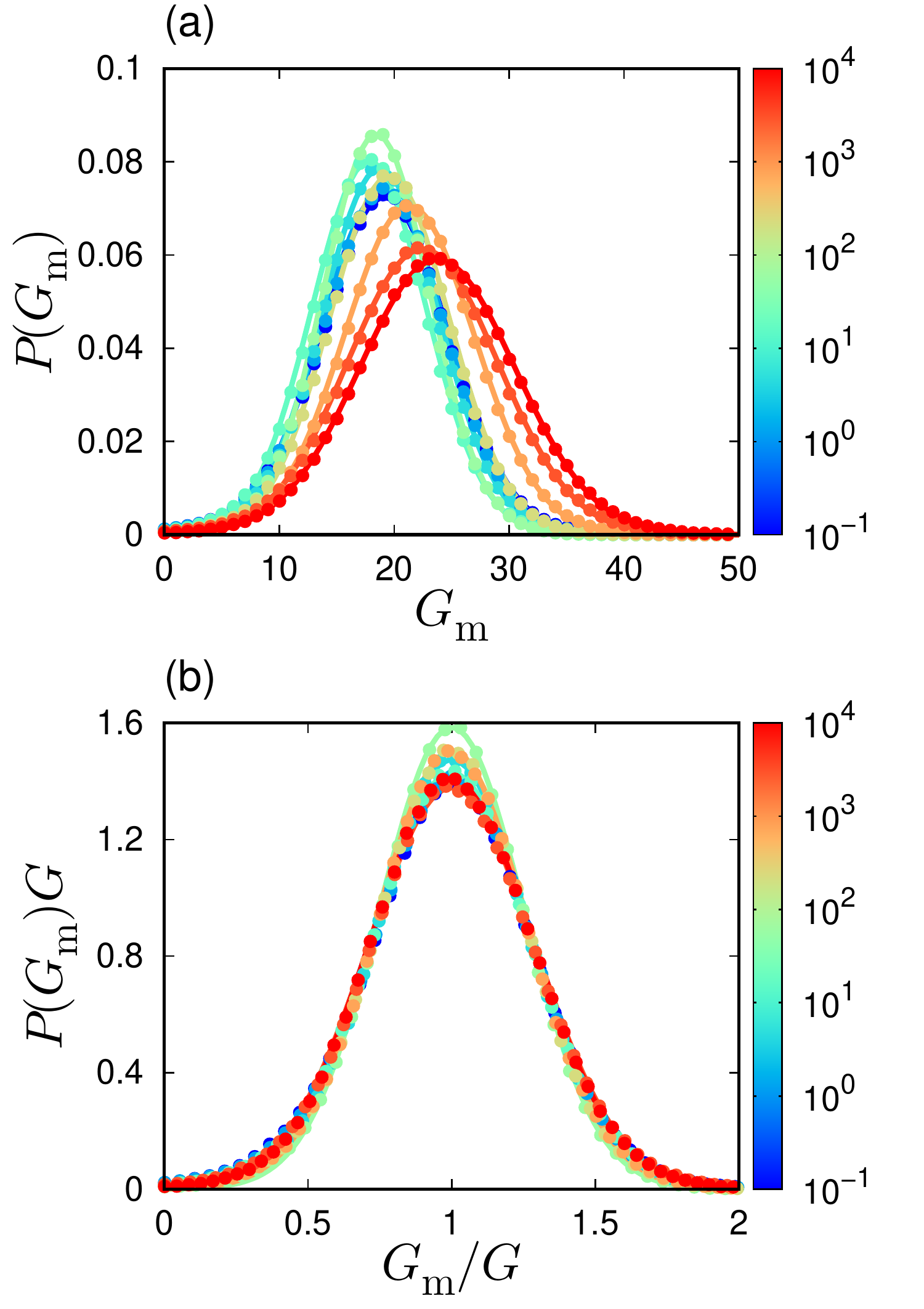}
\vspace*{0mm}
\caption{\label{fig3}
(a) Probability distribution of local shear modulus $P(G_\mathrm{m})$.
The color of the line indicates the value of the bending energy $\varepsilon_\mathrm{bend}$ according to the color bar.
(b) Scaled distribution $P(G_\mathrm{m})G$ as a function of the scaled local shear modulus $G_\mathrm{m}/G$.
The straight lines represent the Gaussian distribution functions fitted to each distribution.
}
\end{figure}

\begin{figure}[t]
\centering
\includegraphics[width=0.48\textwidth]{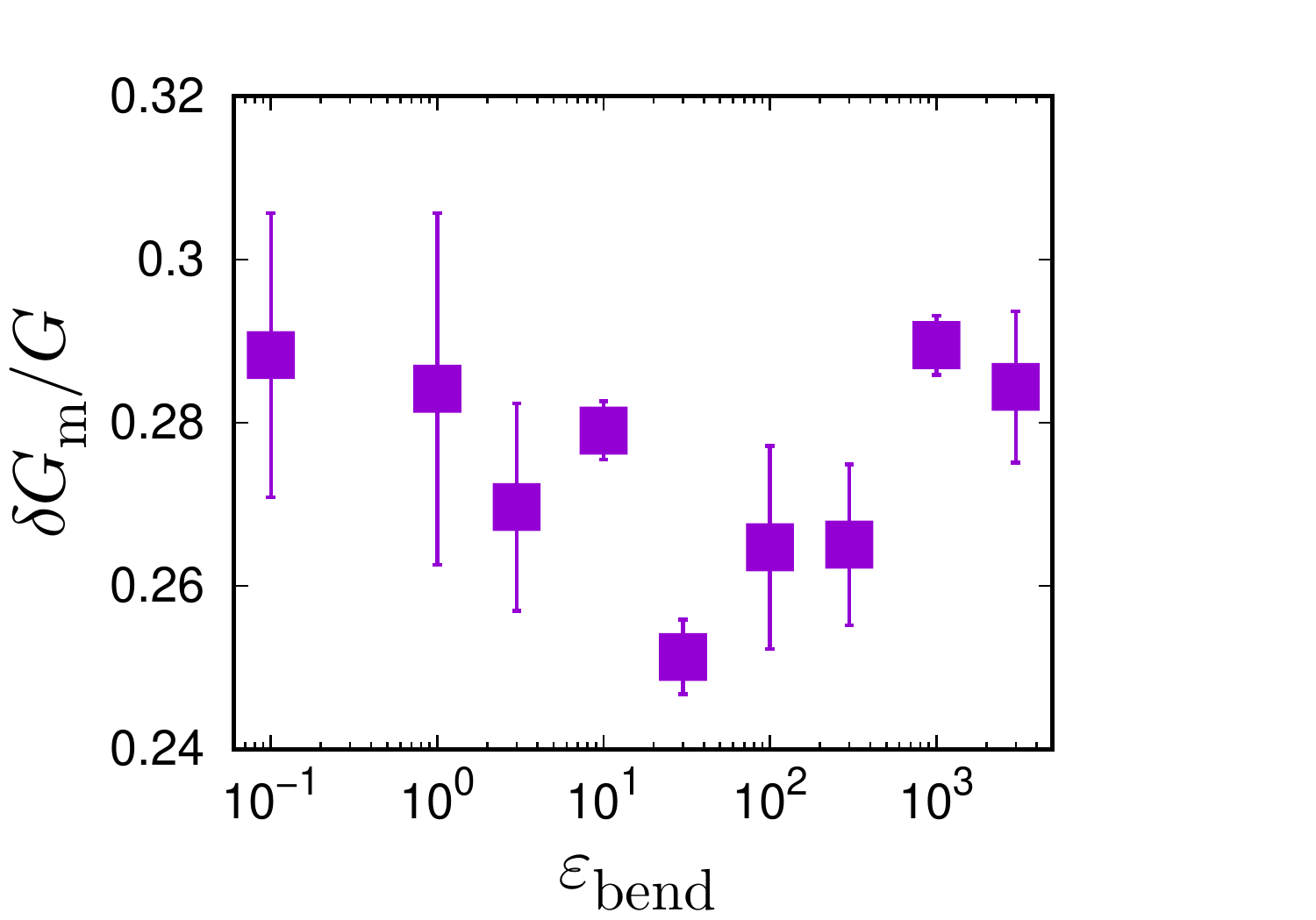}
\vspace*{0mm}
\caption{\label{fig4}
Shear modulus heterogeneity $\delta G_\mathrm{m}/G$ versus bending energy $\varepsilon_\mathrm{bend}$.
}
\end{figure}

\subsection{Local shear modulus distribution}
As demonstrated in Fig.~\ref{fig1}, the reduced vDOS $g(\omega)/\omega^2$ in the BP frequency
$\omega_\mathrm{BP}$ regime was well scaled by using the
Debye frequency $\omega_\mathrm{D}$ and Debye level
$A_\mathrm{D}=3/{\omega_\mathrm{D}}^3$
This suggests that the frequency
and intensity of BP are controlled only by the global sear modulus $G$.
In particular, we obtain the relationship of $\omega_\mathrm{BP}\propto
\omega_\mathrm{D}\propto \sqrt{G}$.
According to the heterogeneous elasticity
theory~\cite{Schirmacher:2006ky, Schirmacher:2007cj, Kohler:2013jj,
Schirmacher:2015bq}, this observation implies that the degree of the shear
modulus heterogeneity $\delta G_\mathrm{m}/G$ is invariant with changes
in the bending energy $\varepsilon_\mathrm{bend}$: this implication is
confirmed below.

We plot the probability distribution of the local shear modulus
$G_\mathrm{m}$ in Fig.~\ref{fig3}(a); this plot follows the Gaussian form
of Eq.~(\ref{eq:Gaussian}).
Figure~\ref{fig3}(b) then plots the scaled distribution
$P(G_\mathrm{m})G$ as a function of the scaled local shear modulus
$G_\mathrm{m}/G$, demonstrating the data of $P(G_\mathrm{m})G$
versus $G_\mathrm{m}/G$ nicely collapse for different values of
$\varepsilon_\mathrm{bend}$.
Because we can transform $P(G_\mathrm{m})$ (Gaussian form) to
\begin{equation}
  P(G_\mathrm{m})G=\frac{1}{\sqrt{2\pi}\left(\frac{\delta G_\mathrm{m}}{G}\right)} \exp{\left[-\frac{\left(\frac{G_\mathrm{m}}{G}-1\right)^2}{2\left(\frac{\delta G_\mathrm{m}}{G}\right)^2}\right]},
\end{equation}
this collapse indicates that the scaled standard deviation $\delta
G_\mathrm{m}/G$ remains unchanged for different
$\varepsilon_\mathrm{bend}$ values.
This is verified by direct demonstration in Fig.~\ref{fig4}, where
$\delta G_\mathrm{m}/G$ is plotted explicitly as a function of
$\varepsilon_\mathrm{bend}$.
Therefore, we can conclude that the bending rigidity of the polymer chain
does not alter the degree of the shear modulus heterogeneity.
This conclusion justifies the theoretical
prediction~\cite{Schirmacher:2006ky, Schirmacher:2007cj, Kohler:2013jj,
Schirmacher:2015bq} 
that vibrational excitations including the BP are controlled only by the
global elastic modulus under the condition of constant heterogeneities
in the moduli distributions.


\begin{figure}[t]
\centering
\includegraphics[width=0.48\textwidth]{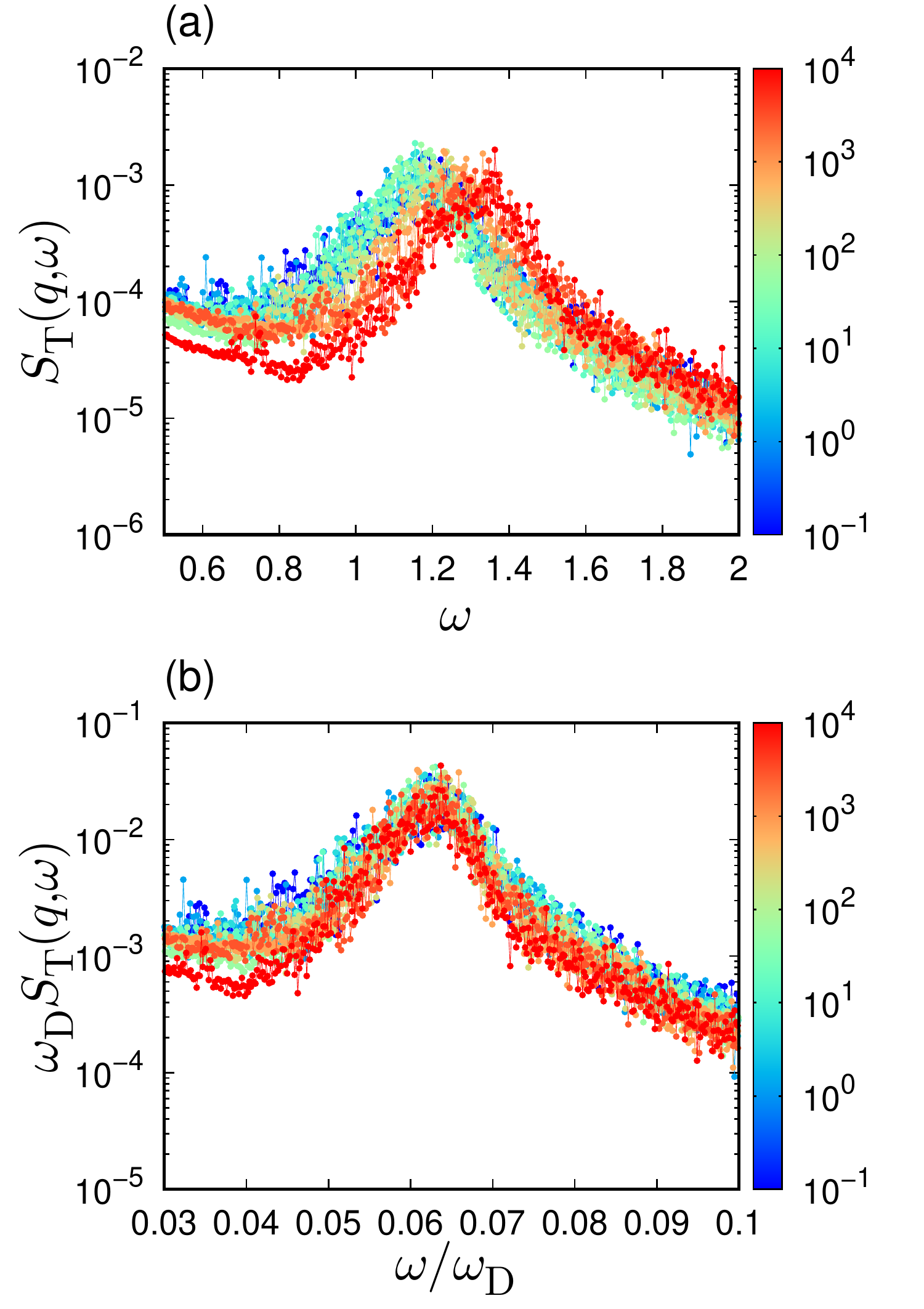}
\vspace*{0mm}
\caption{\label{fig5}
(a) Transverse dynamic structure factor $S_\mathrm{T}(q, \omega)$ as a
 function of $\omega$, at the lowest wave number $q_\mathrm{min}$.
(b) Scaled plot of $S_\mathrm{T}(q, \omega) \omega_\mathrm{D}$ versus $\omega/\omega_\mathrm{D}$.
The color of the line indicates the value of the bending energy $\varepsilon_\mathrm{bend}$ according to the color bar.
}
\end{figure}

\begin{figure}[t]
\centering
\includegraphics[width=0.48\textwidth]{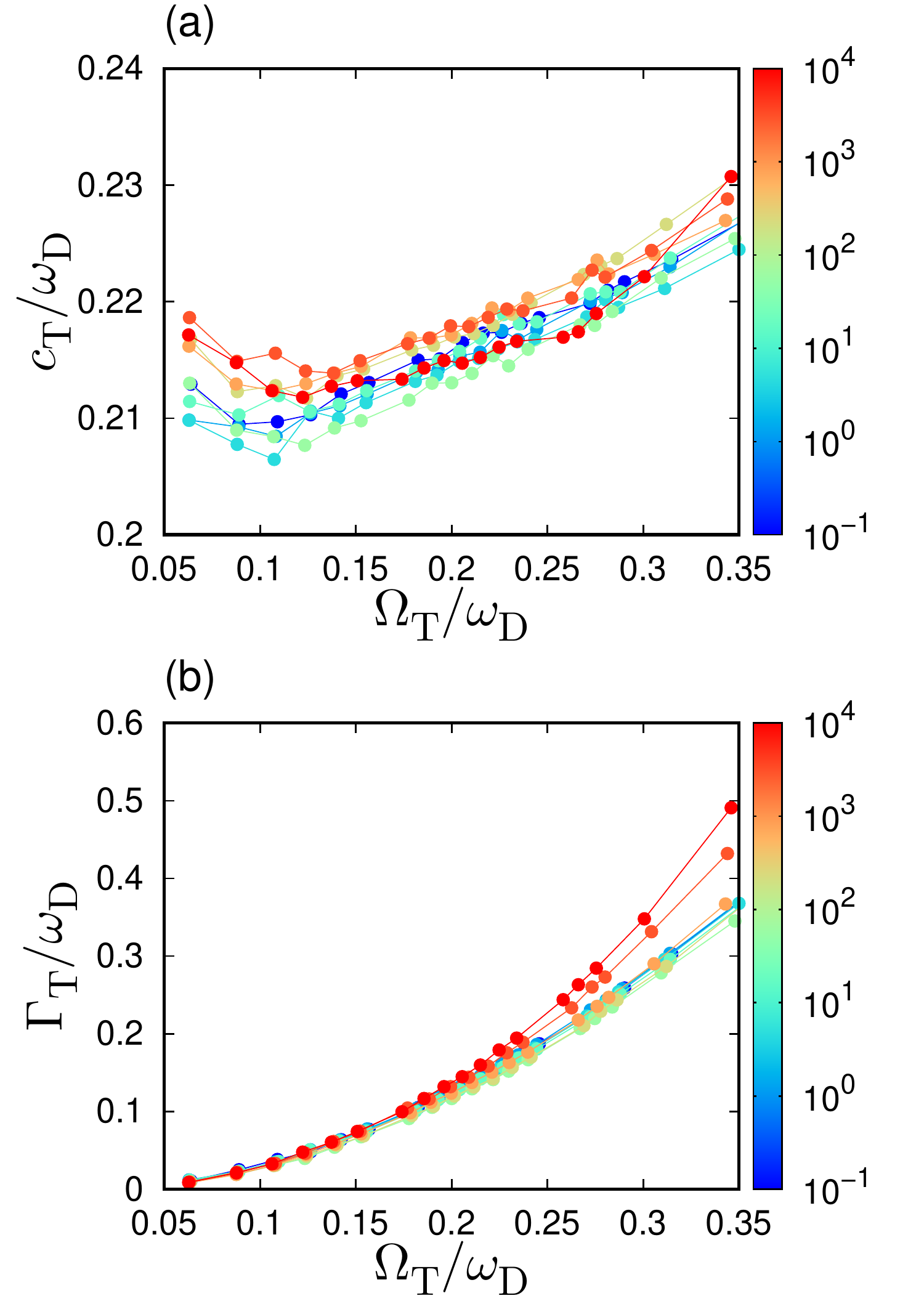}
\vspace*{0mm}
\caption{\label{fig6}
(a) Transverse sound velocity scaled by the Debye frequency,
 $c_\mathrm{T}/\omega_\mathrm{D}$, versus scaled frequency
 $\Omega_\mathrm{T}/\omega_\mathrm{D}$.
(b) Transverse sound damping scaled by the Debye frequency,
 $\Gamma_\mathrm{T}/\omega_\mathrm{D}$, versus
 $\Omega_\mathrm{T}/\omega_\mathrm{D}$.
The color of the line indicates the value of bending energy
 $\varepsilon_\mathrm{bend}$ according to the color bar.
}
\end{figure}

\subsection{Transverse acoustic excitation and its link with boson peak}
We finally study the transverse acoustic excitation in the frequency
regime including the BP.
The generalised Debye model~\cite{Marruzzo:2013ew, Mizuno:2018js} yields the
reduced vDOS $g(\omega)/\omega^2$ in terms of the propagation frequency
$\Omega_\mathrm{T}(q)$ and the attenuation rate $\Gamma_\mathrm{T}(q)$,
as follows:
\begin{equation}
\frac{g(\omega)}{\omega^2} = \frac{3}{\omega_\mathrm{D}^3} + \frac{4}{\pi q_\mathrm{D}^2 c_\mathrm{T}^2(q)} \left[ \frac{\Gamma_\mathrm{T}(q)}{\omega^2+\Gamma_\mathrm{T}^2(q)} \right],
\end{equation}
with Debye wavenumber $q_\mathrm{D}=(6\pi^2\hat\rho)^{1/3}$.
This form can be scaled by $\omega_\mathrm{D}$ and $A_\mathrm{D} = 3/\omega_\mathrm{D}^3$ as:
\begin{equation}
\frac{g(\omega)}{\omega^2 A_\mathrm{D}} = 1 + \frac{4}{3\pi q_\mathrm{D}^2\left(\frac{c_\mathrm{T}(q)}{\omega_\mathrm{D}}\right)^2} \left[ \frac{\left(\frac{\Gamma_\mathrm{T}(q)}{\omega_\mathrm{D}}\right)}{\left(\frac{\omega}{\omega_\mathrm{D}}\right)^2+\left(\frac{\Gamma_\mathrm{T}(q)}{\omega_\mathrm{D}}\right)^2} \right].
\label{eq:reduced_vdos}
\end{equation}
Thus, the collapse of the reduced vDOSs $g(\omega)/(\omega^2 A_\mathrm{D})$
for different values of $\varepsilon_\mathrm{bend}$ indicates that
$c_\mathrm{T}/\omega_\mathrm{D}$ and
$\Gamma_\mathrm{T}/\omega_\mathrm{D}$ are both independent of the bending
energy $\varepsilon_\mathrm{bend}$.

In addition, Eq.~(\ref{eq:DHO}), which is the damped harmonic oscillator
function for the dynamic structure factor $S_\mathrm{T}(q, \omega)$, can
be scaled by the Debye frequency $\omega_\mathrm{D}$:
\begin{equation}
S_\mathrm{T}(q,\omega) \omega_\mathrm{D} \propto
 \frac{\left(\frac{\Gamma_\mathrm{T}(q)}{\omega_\mathrm{D}}\right)\left(\frac{c_\mathrm{T}(q)}{\omega_\mathrm{D}}\right)^2q^2}
 {\left[\left(\frac{\omega}{\omega_\mathrm{D}}\right)^2-\left(\frac{c_\mathrm{T}(q)}{\omega_\mathrm{D}}\right)^2q^2\right]^2+\left(\frac{\omega}{\omega_\mathrm{D}}\right)^2\left(\frac{\Gamma_\mathrm{T}(q)}{\omega_\mathrm{D}}\right)^2},
\label{eq:sqscale}
\end{equation}
which indicates that $S_\mathrm{T}(q,\omega) \omega_\mathrm{D}$ is
simply scaled by $\omega/\omega_\mathrm{D}$, when
$c_\mathrm{T}/\omega_\mathrm{D}$ and
$\Gamma_\mathrm{T}/\omega_\mathrm{D}$ are independent of
$\varepsilon_\mathrm{bend}$.
Below we show that these properties of transverse acoustic excitations are true.

Figure~\ref{fig5}(a) shows the $S_\mathrm{T}(q, \omega)$ for different
values of $\varepsilon_\mathrm{bend}$.
The wave number $q$ is set to its lowest value
$q_\mathrm{min}=2\pi(\hat\rho/N)^{1/3}$, which ranges from
$q_\mathrm{min}=0.283$ (for $\varepsilon_\mathrm{bend}=0.1$) to $0.295$
(for $\varepsilon_\mathrm{bend}=3\times 10^3$).
The frequency of the Brillouin peak shifts to higher values with
increasing $\varepsilon_\mathrm{bend}$.
We then plot $S_\mathrm{T}(q, \omega) \omega_\mathrm{D}$ versus
$\omega/\omega_\mathrm{D}$ in Fig.~\ref{fig5}(b).
It is evident that our calculations of $S_\mathrm{T}(q, \omega)$ are in
accordance with the predicted scaling description of Eq.~(\ref{eq:sqscale}).

We also show the sound velocity $c_\mathrm{T}$ and attenuation rate
$\Gamma_\mathrm{T}$ as functions of the frequency $\Omega_\mathrm{T}$ in
Fig.~\ref{fig6}.
As expected from the scaling property of $g(\omega)$, 
the data of $c_\mathrm{T}$ and $\Gamma_\mathrm{T}$ collapse for different values of
$\varepsilon_\mathrm{bend}$, 
although small deviations are detected.
These collapses are also consistent with the prediction from
Eq.~(\ref{eq:reduced_vdos}) and are explained in terms of the shear modulus heterogeneity.
The collapses break down in the high frequency
regime above the BP frequency, $\Omega_\mathrm{T}/\omega_\mathrm{D}
\gtrsim 0.2 > \omega_\mathrm{BP}/\omega_\mathrm{D} \approx 0.1$.
Because the generalized Debye model does not hold above the BP
frequency~\cite{Marruzzo:2013ew, Mizuno:2018js}, this deviation is not unexpected.

\section{Conclusion}
\label{sec:conclusion}
In conclusion, we have numerically studied elastic heterogeneities and
acoustic excitations in polymer glasses, with particular attention to
the effects of the bending rigidity of the constituent polymer chains.
Our main finding is that the degree of heterogeneity in the local shear modulus
distribution is insensitive to changes in the bending rigidity.
According to the heterogeneous elasticity theory, for unchanging elastic
heterogeneities, the vibrational and acoustic properties of amorphous
materials are controlled only by global elastic moduli.
Consistent with this theoretical prediction, we demonstrated that the
BP and properties of the transverse acoustic excitations are both simply
scaled only by the global shear modulus.
The present work therefore clarified remarkably simple material
property relationships in polymer glasses.
These originate from 
the invariance of the local elastic heterogeneities over an extremely
wide range of bending rigidity values for polymer chains.
Our results also provide good demonstrations that verify the heterogeneous
elasticity theory~\cite{Schirmacher:2006ky, Schirmacher:2007cj,
Kohler:2013jj, Schirmacher:2015bq}, which is among the central
theories used to describe the mechanical and vibrational properties of amorphous
materials.

We note that effects of polymerization on vibrational
properties can be scaled by global elastic moduli~\cite{Caponi:2011ip, Corezzi:2020ha}.
On the contrary, 
some experiments demonstrate that 
the pressure-induced shift of BP cannot be explained by the global
elastic moduli~\cite{Niss:2007de, Hong:2008ix}.
From these observations, we speculate that the polymerization effect is
insensitive to the elastic heterogeneities as is the bending rigidity,
whereas the heterogeneities would be altered by the densification.
Furthermore, recent MD simulations revealed antiplasticizer additives significantly
modify the local elastic constant distribution in
glass-forming polymer liquids~\cite{Riggleman:2010gd}.
It could be interesting to study how boson peak properties change with
evolution of elastic heterogeneities during the antiplasticization process.

At the end of this paper, we would discuss the relationship between the
structural relaxation time and the elastic properties.
Remarkably, numerical work~\cite{Larini:2008gu} has proposed and
demonstrated a scaling relationship between the structural relaxation
time $\tau_\alpha$ and the Debye--Waller factor $\langle u^2\rangle $ for
many types of glass-forming systems, including polymer glasses, as
$\tau_\alpha \propto \exp\left( a \langle u^2\rangle ^{-1} + b \langle
u^2\rangle ^{-2}\right)$ (where $a$ and $b$ are constants).
Because the Debye--Waller factor in the harmonic approximation limit is
estimated as $\langle u^2\rangle  = 3T \int_0^\infty
{g(\omega)}/{\omega^2} d\omega \propto T {\omega_\mathrm{BP}}^{-2} \propto
T G^{-1}$~(where $\omega_\mathrm{BP} \propto \sqrt{G}$ is
applied)~\cite{Shiba:2016bi}, we obtain
\begin{equation} \label{shoving}
\tau_\alpha \propto \exp\left( \alpha \frac{{\omega_\mathrm{BP}}^{2}}{T} +
			 \beta \frac{{\omega_\mathrm{BP}}^{4}}{T^2}
			\right) \propto \exp\left( \alpha' \frac{G}{T} +
			\beta' \frac{G^{2}}{T^2} \right),
\end{equation}
where $\alpha$, $\beta$, $\alpha'$, and $\beta'$ are constants.
This is the idea of the shoving model~\cite{Dyre:1998jm, Dyre:2006im}, which
characterises the activation energy in terms of the global shear modulus
$G$.
Interestingly, Eq.~(\ref{shoving}) has been well demonstrated for
polymer glasses by MD simulations,
where the plateau modulus $G_\mathrm{p}$ of the stress
correlation function was effectively utilized as the shear modulus~\cite{Puosi:2012fx}.
Our results suggest an important condition under which Eq.~(\ref{shoving})
holds.
When the spatial heterogeneity in the local shear modulus
distribution is 
unchanged, the excess vibrational excitations, \textit{i.e.}, the BP, are
controlled only by the global shear modulus, indicating that the structural
relaxation time is also controlled solely by the global shear modulus.

\begin{acknowledgments}
This work was supported by JSPS KAKENHI Grant Numbers:
JP19K14670~(H.M.), JP20H01868~(H.M.), JP18H01188~(K.K.),
JP20H05221~(K.K.), and JP19H04206~(N.M.).
This work was also partially supported by the Asahi Glass Foundation and
by the Fugaku Supercomputing Project and the Elements Strategy
Initiative for Catalysts and Batteries (No.~JPMXP0112101003) from the
Ministry of Education, Culture, Sports, Science, and Technology.
The numerical calculations were performed at Research Center of
Computational Science, Okazaki Research Facilities, National Institutes
of Natural Sciences, Japan.
\end{acknowledgments}

%

\end{document}